\documentstyle[12pt]{article}
\begin{document}

\centerline{\Large Folded Bifurcation in Coupled Asymmetric Logistic Maps}

\vskip 3cm

\centerline{Shoichi MIDORIKAWA}
\centerline{Faculty of Engineering, Aomori University, 2-3-1 Kobata, 
Aomori 030, Japan}

\vskip 1cm

\centerline{Takayuki KUBO}
\centerline{Institute for Nuclear Study, University of Tokyo, Tanashi, 
Tokyo 188, Japan}

\vskip 1cm

\centerline{Taksu CHEON}
\centerline{Department of Physics, Hosei University, Chiyoda-ku, Fujimi, 
Tokyo 102, Japan}

\vskip 5cm

\centerline{\bf Abstract}

\vskip 5mm

A system of coupled two logistic maps, one periodic and the other chaotic, 
is studied. It is found that with the variation of the coupling strength, 
the system displays several curious features such as the appearance of 
quadrupling of period, occurrence of isolated period three attractor and 
the coexistence of the Hopf and pitchfork bifurcations. Possible applications 
and extensions are discussed.

\newpage

The chaos in higher dimensional system is one of the focal subject of 
physics today. Along with the approach starting from modeling physical 
system with many degrees of freedom, there emerged a new approach, developed 
by Kaneko to couple many one dimensional maps [1-3] to study the behavior 
of the system as a whole. However, this model can only be applied to study 
the dynamics of a single medium such as the pattern formation in a fluid. 
What happens if two media border on each other? One may naturally leads 
to the model of coupled logistic maps with different strength parameters. 
Thus it is appropriate to inquire whether loosening the condition of strict 
identity might bring any new feature while keeping both maps to be logistic 
to hold the redundant controlling parameters minimal.  
Even two logistic maps coupled to each other may serve as the dynamical model 
of driven coupled oscillators. It has been found that two coupled identical 
maps possess several characteristic features which are typical of higher 
dimensional chaos [1,4]. 
 
In this letter, we report the results of numerical investigation on the 
system of two logistic maps with different strength parameters such that 
the one map lies in a period one stable attractor or a bifurcation point 
and the other in chaotic region when decoupled. Several new features 
previously unobserved are found. Most notable among them is the appearance 
of a period four cycle straight from the stable period one cycle. The other 
peculiar feature is the almost simultaneous occurrence of periods four, 
eight and sixteen right after the Hopf bifurcation which results in a very 
intriguing metamorphose of the attractor when one changes the coupling 
parameter.

The system we study is two linearly coupled maps
\begin{eqnarray}
x_{n+1} &=& (1 - \epsilon) f(\mu,x_n) + \epsilon f(\nu,y_n) \nonumber \\
y_{n+1} &=& \epsilon f(\mu,x_n) + (1 - \epsilon) f(\nu,y_n)
\end{eqnarray}
where the map $f$ is taken to be the logistic map [5] with strength parameters 
$\mu$. 
\begin{eqnarray}
f(\mu,x) = \mu x (1 - x)
\end{eqnarray}
In case of ${\mu=\nu}$, two maps soon become synchronized no matter 
what initial conditions may be, {\it i.e.}, coupled maps 
are identical with a single logistic map. The interesting is the case of 
${\mu \ne \nu}$. In the following    
we fix the strength parameters above and below critical value 
${\mu^* = 3.56994 \ldots}$ for $\mu$ and $\nu$ respectively. We choose the 
strength parameters $\mu$ and $\nu$ and regard the coupling parameter 
$\epsilon$ as the controlling parameter.

In Fig. 1, the attractors of the coupled-map are displayed as functions of 
coupling $\epsilon$. Fig. 1(a) shows the result of ${\mu=4}$ and ${\nu=3}$, 
and (b) that of ${\mu=4}$ and ${\nu=2}$. They are two typical example of 
various values of $\mu$ and $\nu$. One immediately notices several interesting 
features. The fact that there are two chaotic regions in both ${\epsilon=0}$ 
and ${\epsilon=1}$ ends seems odd at first sight, but after some reflection, 
one realizes that very weak $\epsilon$ means very strong ${(1 - \epsilon)}$ 
which brings chaos first to the variable $x$, then to $y$ however weak the 
coupling term may be. 
The most salient feature is the appearance of stable period 
four cycle right after the period one around ${\epsilon=0.77}$ in Fig. 1(a). 
Another, found in both (a) and (b) cases, is the sudden filling of the $x$ 
and $y$ space around ${\epsilon=0.85}$ and above. The broad window-like region 
with period four around ${\epsilon=0.9}$ in the case of (b) is also 
noteworthy. 
 
Identifying stable and unstable periodic points often gives the skeleton view 
of the dynamics of the system. With the vector notation of the variables 
${\vec{z}=(x,y)}$ and the operator notation of the map 
\begin{eqnarray}
\vec{z}_{n+1} = \vec{F}(\vec{z}_n) , 
\end{eqnarray}
one can define the periodic points of cycle $N$ as the solution of the equation
\begin{eqnarray}
\vec{z}_N = \vec{F}^N(\vec{z}_N) , 
\end{eqnarray}
where ${\vec{F}^N(\vec{z}_N)}$ is defined by 
${\vec{F}^N(\vec{z}_N) = \overbrace{\vec{F}(\vec{F}(\vec{F} \cdots \vec{F}
(}^{N \ times}\vec{z}_N) \cdots))}$.     
Bifurcations occur at the parameter value of $\epsilon$ where the periodic 
point satisfies 
\begin{eqnarray}
\left| E_{max} \left[\frac{d\vec{F}^N}{d\vec{z}} (\vec{z}_N) \right] \right| 
= 1 ,
\end{eqnarray}
in addition to eq.(4). Here, ${E_{max}[G]}$ is the maximum eigenvalue of the 
matrix $G$. We numerically solve eq.(4) for ${N=3,4,}$ and $8$ for entire 
range of ${\epsilon=0 \sim 1}$. The results are presented in Fig. 2 where 
the unstable orbits are shown as well as the stable ones. One can clearly 
see from ${N=4}$ and ${N=8}$ skeleton, that the period four occurring at 
${\epsilon=0.77}$ is the result of the {\it folded period doubling bifurcation 
sequence} whose period two starts at ${\epsilon=0.75}$ after the quadrupling 
of the period when one moves from period one at larger $\epsilon$ to a higher 
periodic orbits at smaller $\epsilon$. While this type of inversion of the 
sequence is nothing of unimaginable nature in principle, it has never been 
observed in simple maps to our knowledge. One recognizes easily that it is 
rather hard to obtain in a usual one-dimensional map since it requires special 
tuning of functional dependence of the map on the controlling parameter to 
ensure that the condition eq.(4) is satisfied for ${N=4}$ before ${N=2}$. 
We therefore think it significant that it emerges from nothing but linear 
coupling of two logistic maps of different strength parameters. One might 
presume that it should be observed rather frequently in higher dimensional 
coupled lattice map models once the requirement of strict uniformity of the 
elements is lifted. In fact, preliminary investigation on three coupled 
logistic maps supports this view.

There are two notable facts in the periodic cycle diagram above 
${\epsilon=0.85}$. One is that no periodic point exists in the region 
${\epsilon < 0.86}$. The examination of the ${x-y}$ profile shows that the 
onset of ``filling area'' at approximately ${\epsilon=0.85}$ is the result of 
the Hopf bifurcation[6]. The second is that around ${\epsilon=0.88}$, period 
four and period eight start (as well as other higher periods not shown here) 
appearing almost simultaneously. While these occurrences fit into the generic 
transition scenario of 
``${cycle \ one \to Hopf \to}$ ${cycles \to \ldots}$ ${\to chaos}$''  
found by Kaneko in an early study[1,3], the second fact, crowded onset of 
many different periodic cycles makes this transition to chaos a very intricate 
one. These points are visually displayed in the phase portrait of the 
attractor at several values at and above ${\epsilon=0.85}$ shown in Fig. 3. 
The first two figures (a) and (b) clearly show the Hopf bifurcation around 
${\epsilon=0.85}$. Subsequent distortion and transformation to an aurora-like 
strange attractor observed in Fig. 3(c) to (f) certainly match those found in 
far more complicated system in its aesthetic appeal. One curious feature found 
both in Figs. 2 and 3(b) is the early appearance of period 3 cycle around 
${\epsilon=0.87}$. At present, we are unable to notice any direct role of this 
period to the shape and stability of the attractors around this region[7].  

In summary, we have constructed a system consisting of chaotic and periodic 
maps coupled together. While the dynamics of the system does not go beyond the 
known bifurcation schemes in the two-dimensional dissipative system (which is 
certainly not to be expected), it is found that various bifurcations occur in 
such a combination to give the system several intriguing features.      

Finally, few words on possible applications and extensions are in order. It 
should not be too difficult to construct a circuit (electronic, for example) 
to materialize the system proposed here. Actual applicational usage of 
quadrupling of stable states might be envisioned. Also, replacing the logistic 
map with other maps, cycle map or quadratic map for example, might be 
interesting to see the generality and/or the new aspects of the findings here. 
Another application is possible by increasing the number of maps coupled each 
other. The coupled lattice map having ``impurity'', or one element with 
different strength parameter from all the rest might exhibit non-conventional 
features. A lattice with alternating strength might also be an interesting 
system. In a word, loosening the condition of strict uniformity of the 
elements might bing some new feature to the coupled lattice map.
  
\vskip 1cm

We express our gratitude to O. Morimatsu for the supports in computational 
aspects of the work. The numerical calculation has been performed on VAX3100 
of the Theory Division, Institute for Nuclear Study, VAX6000/400 at Nuclear 
Science Center, Department of Physics, University of Tokyo and also on Sparc 
Station 2 at Department of Physics, Hosei University.  

\newpage  

\centerline{\bf References}

\vskip 1cm

\begin{enumerate}
\item K. Kaneko, Prog. Theor. Phys. {\bf 69}, 1427 (1983).
\item K. Kaneko, Prog. Theor. Phys. {\bf 72}, 480 (1984); 
      {\it ibid.} {\bf 74}, 1033 (1985).
\item K. Kaneko, Collapse of tori and genesis of chaos in dissipative systems 
(World Scientific, 1986).
\item T. Hogg and B.A. Huberman, Phys. Rev. A{\bf 29}, 275 (1984).
\item R. May, Nature {\bf 261}, 459 (1976).
\item E. Hopf, Ber. Math.-Phys. K1. S\"achs. Akad. Wiss. (Leipzig), {\bf 94}, 
1 (1942).
\item D. Ruelle and F. Takens, Commun. Math. Phys. {\bf 20}, 167 (1971).
\end{enumerate} 

\newpage 
\centerline{Figure Captions}

\vskip 1cm

\begin{enumerate}
\item[Fig. 1] The phase diagram of the coupled maps eqs.(1)-(2). (a) is 
for the parameter values ${\mu = 4.0}$, ${\nu = 3.0}$, and (b) is for 
${mu = 4.0}$, ${\nu = 2.0}$.
\item[Fig. 2] The ``skelton'' of Fig. 1, showing the position of the 
periodic cycles of the maps. (a) is for ${\mu = 4.0}$, ${\nu = 3.0}$, and 
(b) for ${\mu = 4.0}$, ${\nu = 2.0}$ as before. The caption number indicates 
the period of the cycle. 
\item[Fig. 3] Spatial portraits of the attractors for the coupled logistic 
map with ${\mu = 4.0}$ and ${\nu = 3.0}$ at ${\epsilon = 0.852}$ (a), 
${\epsilon = 0.868}$ (b), ${\epsilon = 0.8757}$ (c), ${\epsilon = 0.877}$ (d), 
${\epsilon = 0.885}$ (e), and ${\epsilon = 0.900}$ (f).

\end{enumerate}

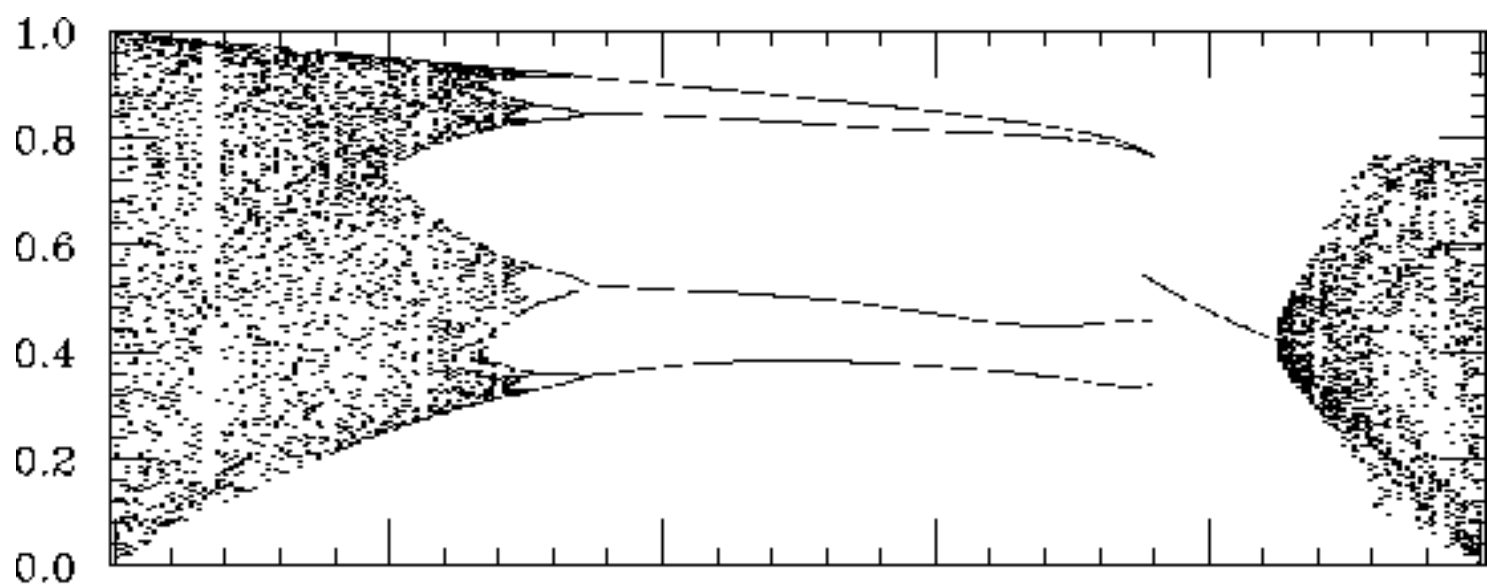
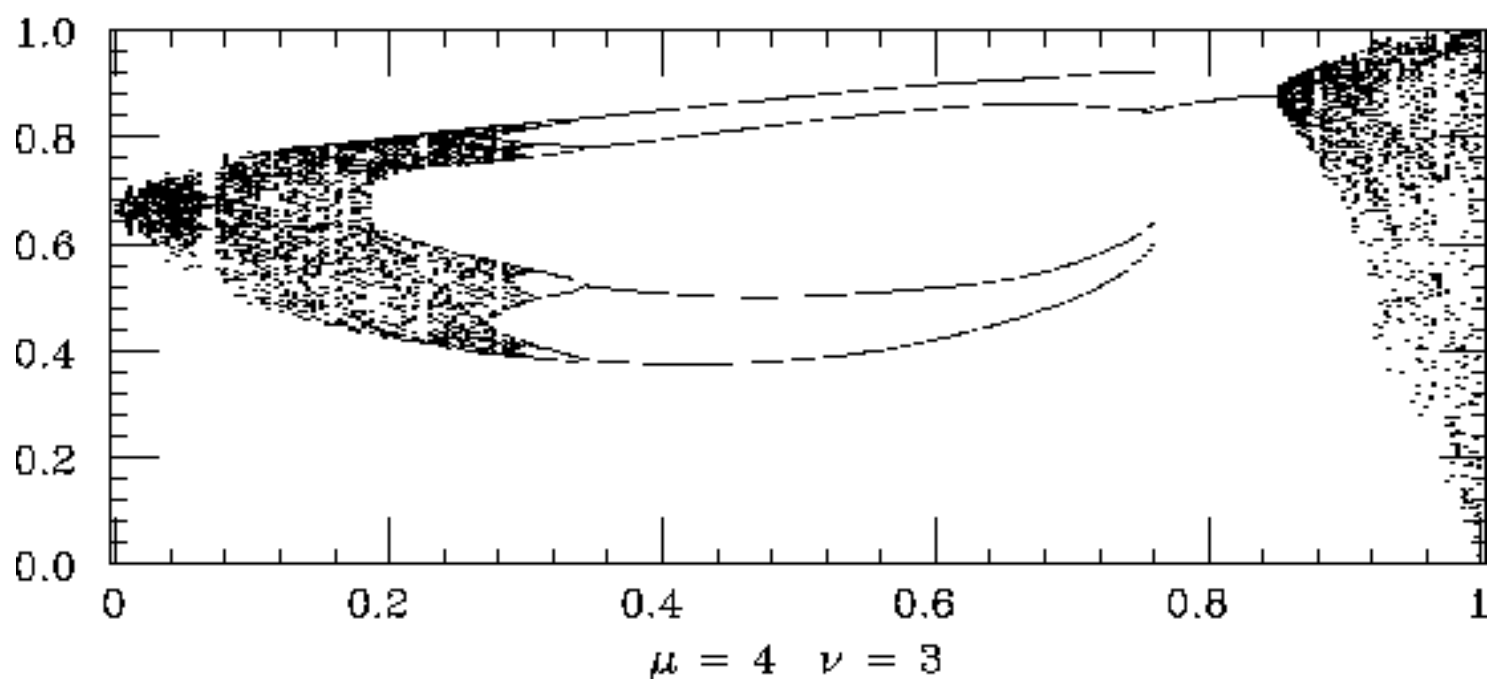
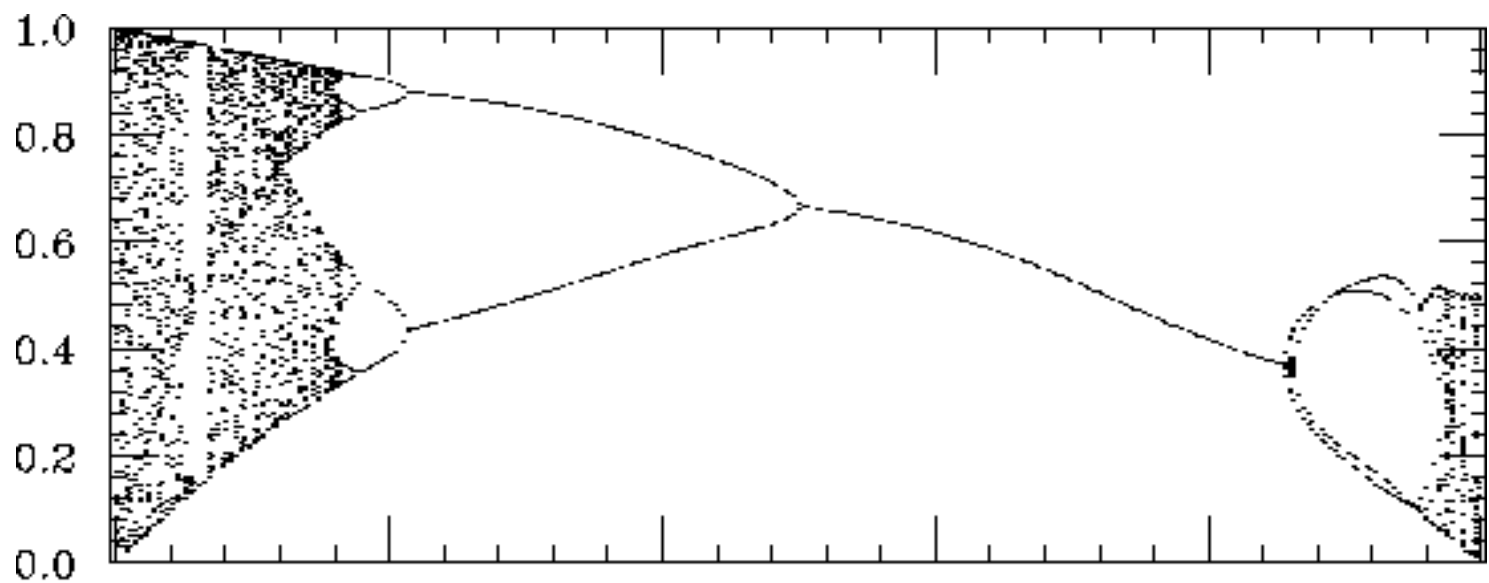
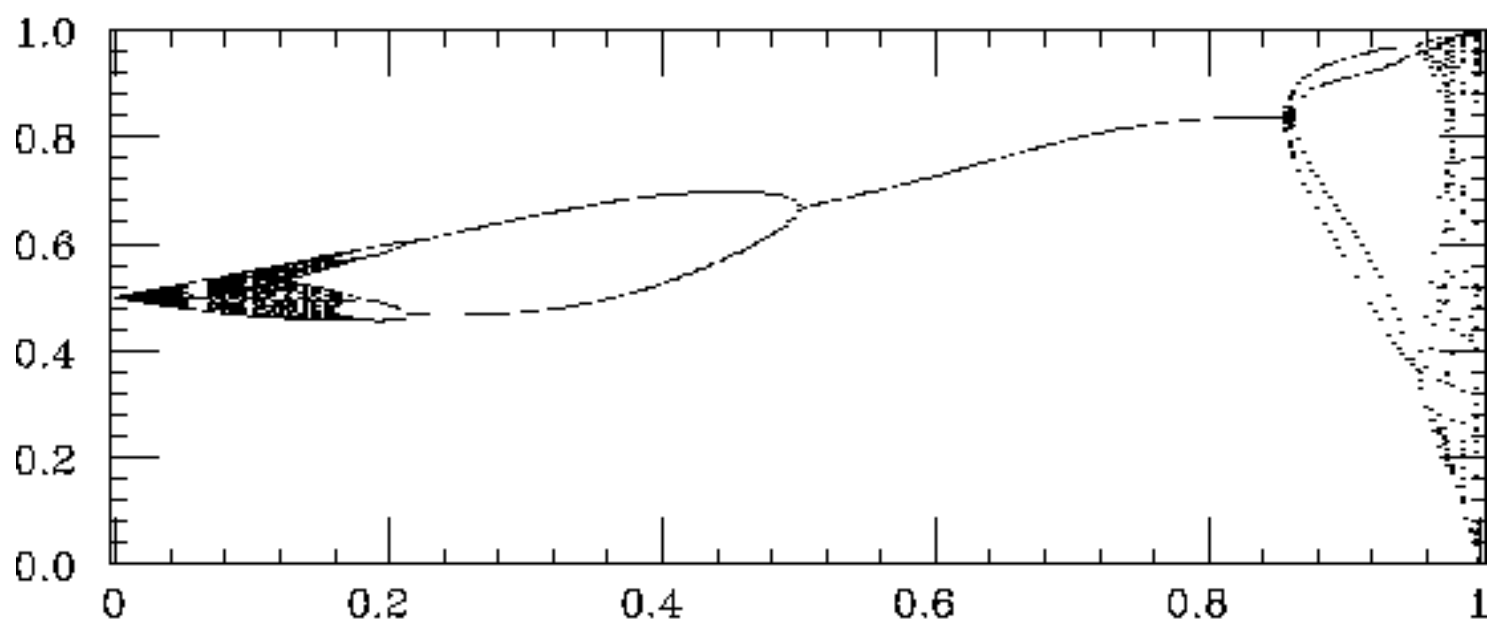
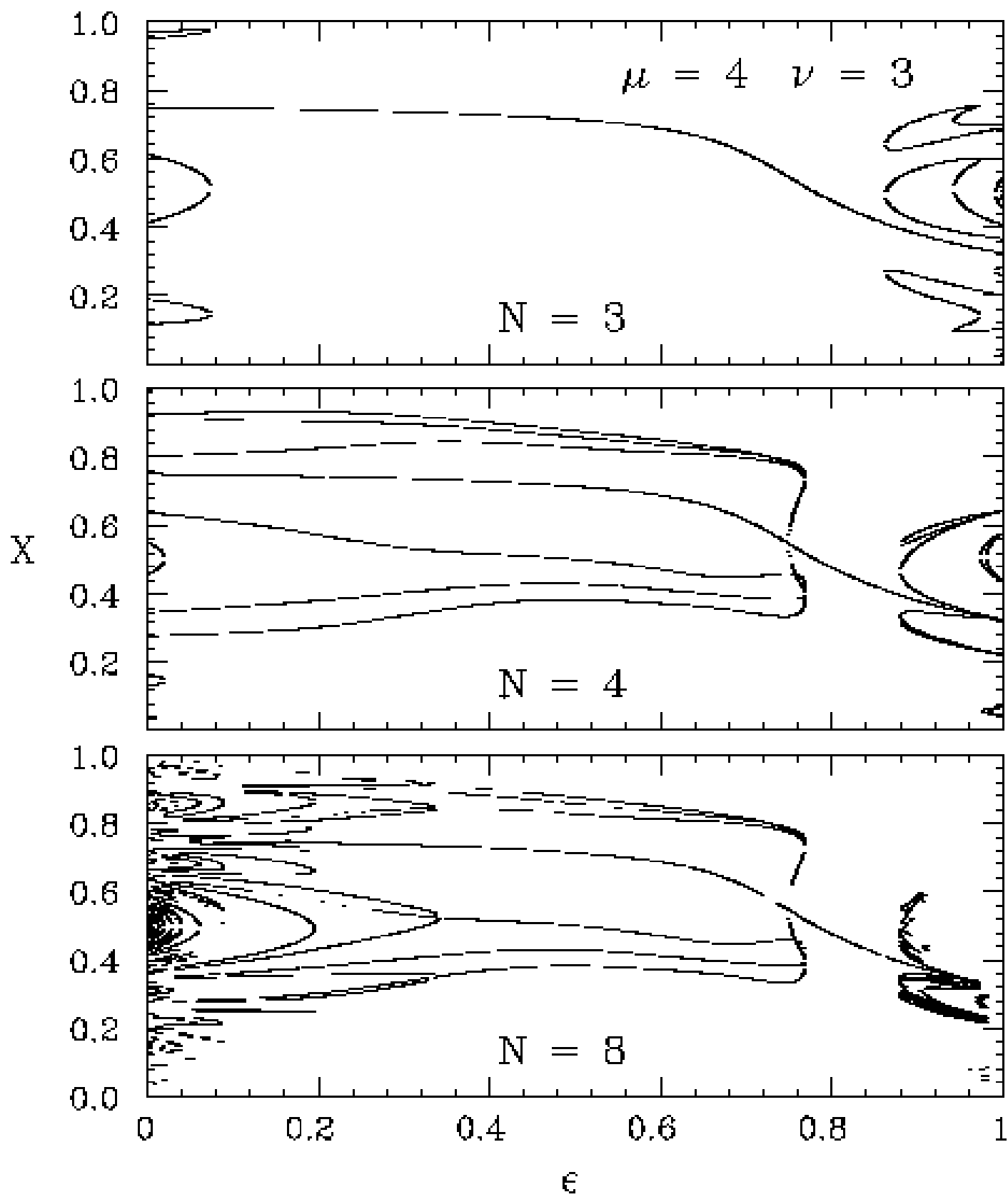
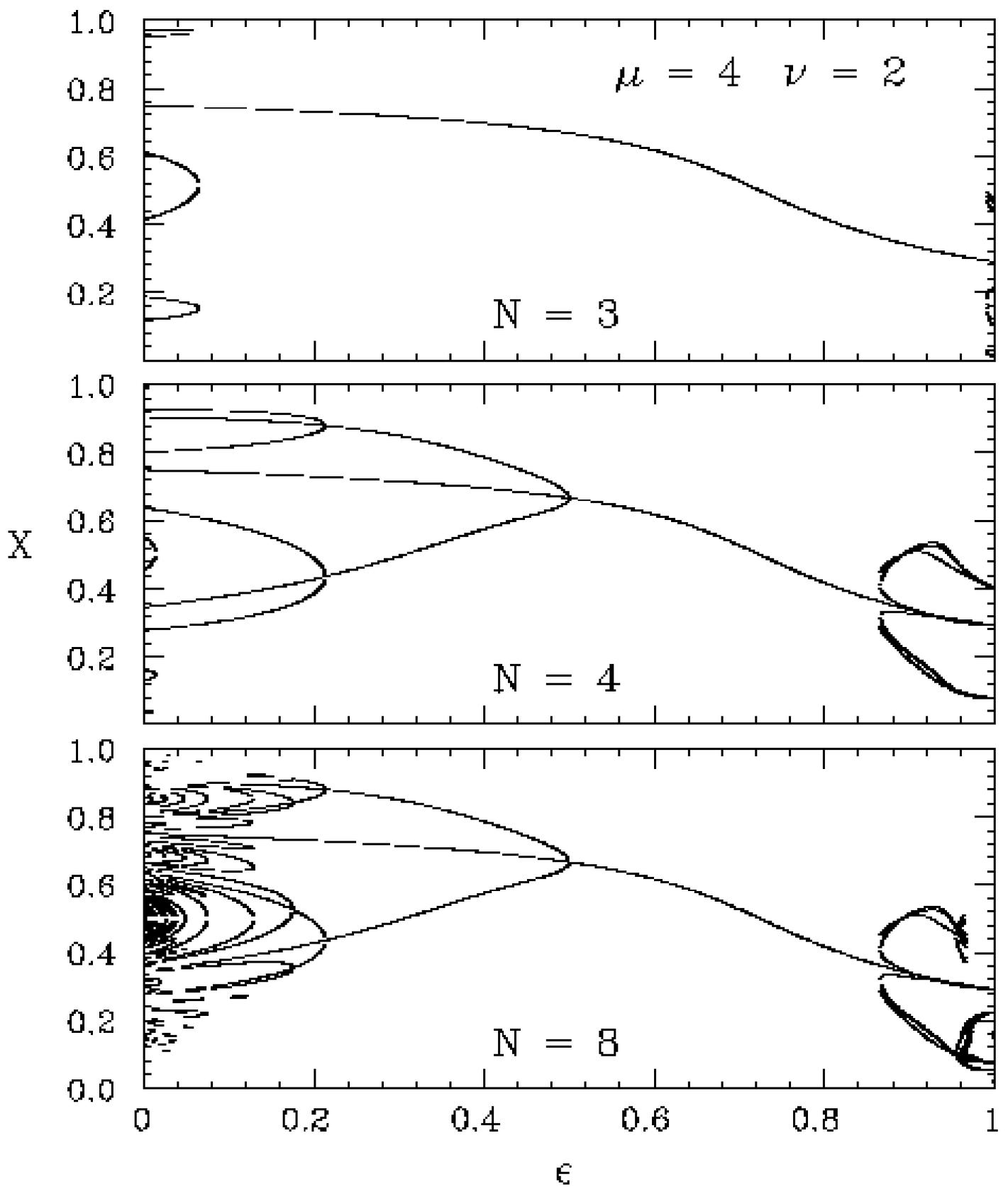
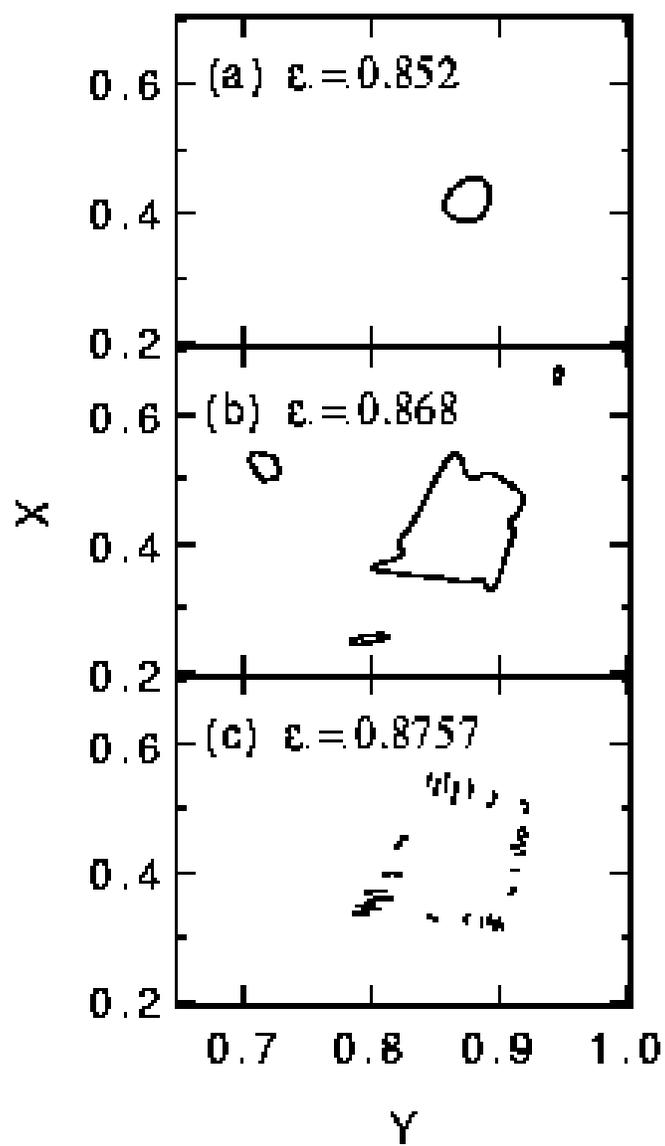
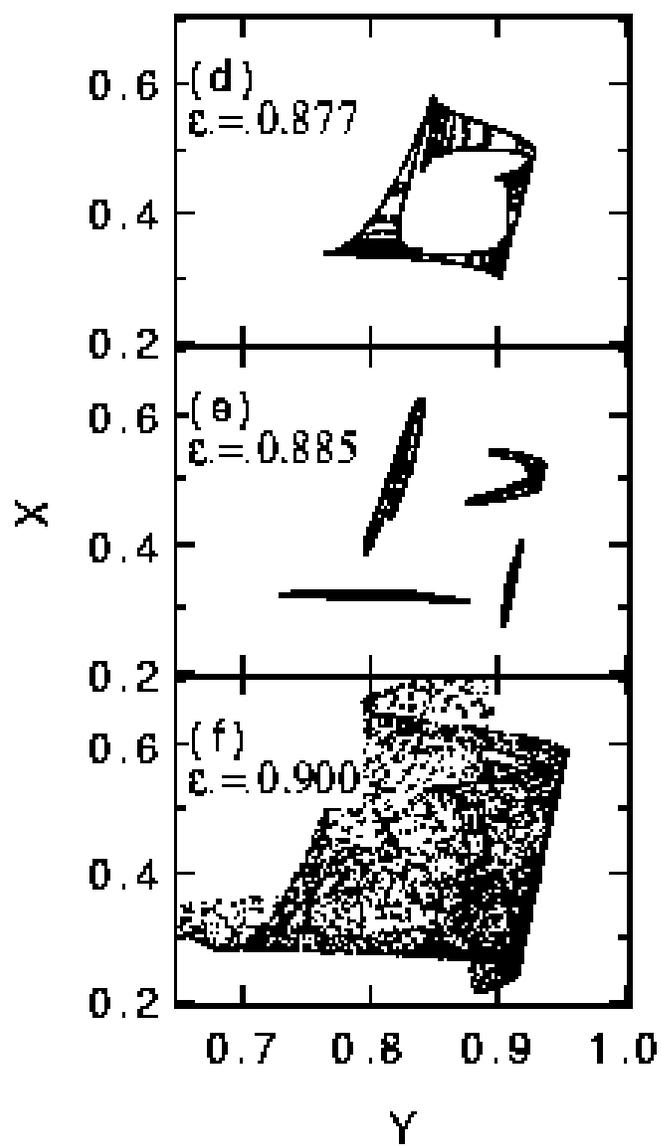
\end{document}